\documentclass[sigconf]{acmart}

\usepackage{enumitem}
\usepackage{rotating}
\usepackage{ccicons}          
\usepackage{soul}

\usepackage{siunitx}

\copyrightyear{2026}
\acmYear{2026}
\setcopyright{cc}
\setcctype{by-nc-nd}
\acmConference[CHI '26]{Proceedings of the 2026 CHI Conference on Human Factors in Computing Systems}{April 13--17, 2026}{Barcelona, Spain}
\acmBooktitle{Proceedings of the 2026 CHI Conference on Human Factors in Computing Systems (CHI '26), April 13--17, 2026, Barcelona, Spain}
\acmPrice{}
\acmDOI{10.1145/3772318.3790845}
\acmISBN{979-8-4007-2278-3/2026/04}

\AtBeginDocument{%
  }

\begin{document}

\title[AI-Mediated Video Communication Reduces Interpersonal Trust and Confidence in Judgments] {Through the Looking-Glass: AI-Mediated Video Communication Reduces Interpersonal Trust and Confidence in Judgments}

\begin{abstract}
AI-based tools that mediate, enhance or generate parts of video communication may interfere with how people evaluate trustworthiness and credibility. In two preregistered online experiments (N = 2,000), we examined whether AI-mediated video retouching, background replacement and avatars affect interpersonal trust, people's ability to detect lies and confidence in their judgments. Participants watched short videos of speakers making truthful or deceptive statements across three conditions with varying levels of AI mediation. We observed that perceived trust and confidence in judgments declined in AI-mediated videos, particularly in settings in which some participants used avatars while others did not. However, participants' actual judgment accuracy remained unchanged, and they were no more inclined to suspect those using AI tools of lying. Our findings provide evidence against concerns that AI mediation undermines people's ability to distinguish truth from lies, and against cue-based accounts of lie detection more generally. They highlight the importance of trustworthy AI mediation tools in contexts where not only truth, but also trust and confidence matter.
\end{abstract}

\author{Nelson Navajas Fernández}
\authornote{Corresponding author e-mail: nelson.navajas.fernandez@uni-weimar.de}
\orcid{0009-0009-4298-9658}
\affiliation{%
  \institution{Bauhaus University}
  \city{Weimar}
  \country{Germany}
}
\author{Jeffrey T. Hancock}
\orcid{0000-0001-5367-2677}
\affiliation{%
  \institution{Stanford University}
  \city{Stanford}
  \country{United States}
}

\author{Maurice Jakesch}
\orcid{0000-0002-2642-3322}
\affiliation{%
  \institution{Bauhaus University}
  \city{Weimar}
  \country{Germany}
}

\ccsdesc[500]{Human-centered computing~Empirical studies in collaborative and social computing}
\ccsdesc[500]{Human-centered computing~Interaction design theory, concepts and paradigms}
\ccsdesc[500]{Computing methodologies~Artificial intelligence}

\keywords{AI-mediated communication, video filters, deception detection, trust, credibility, avatars, experiments}


\maketitle

\section{Introduction}
\begin{figure*}
  \begin{center}
    \includegraphics[width=.9\textwidth]{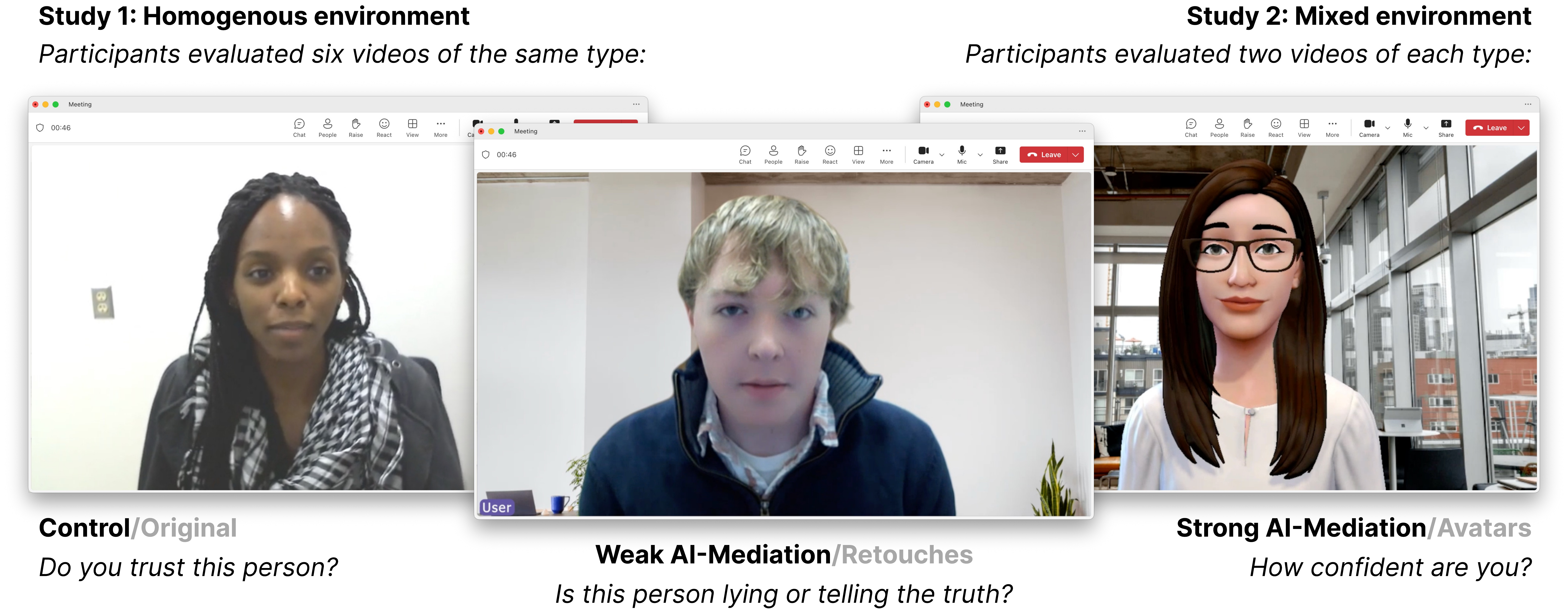}
  \end{center}
\caption{\textbf{Study conditions and primary outcome variables.} Participants watched six videos in which video subjects recounted a story about someone they knew, that was either true or false. In the control condition, we embedded the original video in a video call to increase realism. In the weak AI-mediated treatment, we further processed the video using retouching and a virtual background. In the strong AI-mediated treatment, we replaced the subject with an animated avatar. For each video, participants indicated whether they trusted the person in the video, whether they thought the person in the video was lying, and how confident they were in their judgment.}
\Description{Study conditions and primary outcome variables. Participants watched six videos in which video subjects recounted a story about someone they knew, that was either true or false. In the control condition, we embedded the original video in a video call to increase realism. In the weak AI-mediated treatment, we further processed the video using retouching and a virtual background. In the strong AI-mediated treatment, we replaced the subject with an animated avatar. For each video, participants indicated whether they trusted the person in the video, whether they thought the person in the video was lying, and how confident they were in their judgment.}
\label{fig:stimuli}
\end{figure*}

In 2021, a lawyer joined a virtual court hearing in Texas, to everyone's surprise, appearing as a wide-eyed kitten \cite{bbcnewsLawyerGetsStuck2021}. "I'm here live. I'm not a cat" he clarified and excused himself with a kitty-worried expression, blinking at the judge. When such mishaps make the news, these incidents demonstrate how deeply AI communication systems can disrupt the expectations and assumptions people hold about mediated communication, particularly in high-stakes contexts where what is said has far-reaching consequences.

However, most AI-based transformations we use in our communication today are less obvious than a cat avatar. With the shift to remote communication accelerated by the COVID-19 pandemic \cite{vargoDigitalTechnologyUse2021,shockleyRemoteWorkerCommunication2021}, video communication platforms are increasingly used in not only casual but also professional and high-stakes settings \cite{mannCOVID19TransformsHealth2020,doringVideoconferenceFatigueConceptual2022,oconaillConversationsVideoConferences1993,jacksResearchRemoteWork2021}. Platforms such as Zoom, Google Meet and Microsoft Teams, integrate a wide range of algorithmic video enhancements and transformations. The available features range from background blurring and replacement to skin improvements, gaze correction and personalized avatars that resemble the speaker. 

AI-based video features are widely used and broadly regarded as acceptable \cite{doringVideoconferenceFatigueConceptual2022a,javornikWhatLiesFilter2022,zoomZoomVirtualBackgrounds2023}. While often presented as convenience features or aesthetic improvements \cite{zoomZoomVirtualBackgrounds2023}, they may alter aspects of communication that are central to impression formation and judgment \cite{hancockAIMediatedCommunicationDefinition2020,hohensteinAIMoralCrumple2020,jakeschAIMediatedCommunicationHow2019}. As AI-mediated video tools become widely deployed, we need to better understand their potential to shape people's impressions and judgments \cite{hancockAIMediatedCommunicationDefinition2020,hancockSocialImpactDeepfakes2021} as well as trust \cite{jakeschAIMediatedCommunicationHow2019,maSelfDisclosurePerceivedTrustworthiness2017}, honesty \cite{hohensteinAIMoralCrumple2020,leibCorruptedAlgorithmsHow2024,suenRevealingInfluenceAI2024} and credibility \cite{kimPerceivedCredibilityAI2022} in online video communication.

Indeed, previous research in computer-mediated communication (CMC) shows that the medium of interaction can significantly shape how we present ourselves and how others perceive us \cite{hancockSeeNoEvil2010}. In video communication, many cues that people rely on in face-to-face settings, such as posture, eye contact and microexpressions \cite{connellySignalingTheoryReview2011}, are missing or obscured. Even in traditional computer-mediated communication, judging the honesty, credibility and trustworthiness of others is a difficult task \cite{bondAccuracyDeceptionJudgments2006,depauloCuesDeception2003,levineTruthDefaultTheoryTDT2014}. Because such evaluations shape how people interpret and respond to others in everyday communication \cite{levineTruthDefaultTheoryTDT2014,vrijDetectingLiesDeceit2000}, preserving people's ability to evaluate others in AI-mediated communication remains essential.  

The introduction of AI-based video tools may affect the ways people form impressions of each other and assess credibility, authenticity and trustworthiness \cite{hohensteinAIMoralCrumple2020,jakeschAIMediatedCommunicationHow2019,kimPerceivedCredibilityAI2022}. Hancock et al. conceptualized the relevant changes under the framework of AI-mediated communication (AI-MC) \cite{hancockAIMediatedCommunicationDefinition2020}: a paradigm shift in computer-mediated communication where a computational agent modifies, augments or generates message content on behalf of a communicator. Prior work in text-based contexts has shown that a mediating AI system that modifies or creates communication can erode interpersonal trust \cite{jakeschAIMediatedCommunicationHow2019} and alter how receivers interpret intentions, credibility and agency \cite{gliksonAImediatedApologyMultilingual2023,mieczkowskiAIMediatedCommunicationLanguage2021a,purcellPeopleHaveDifferent2024a}. In video communication---the context of the current study---interactions are more dynamic and perceptually rich than text, further complicating assessments of how the AI tools integrated into widely used platforms \cite{zoomZoomVirtualBackgrounds2023} may affect judgments of trust, honesty and credibility.

This study investigates how different levels of AI-mediated video communication---ranging from original recordings to weak AI mediation through retouching and virtual backgrounds to strong AI mediation through animated avatars---affect the perceived trustworthiness of the speaker, judgments of truth and confidence in these judgments. In two large online experiments (N = 2,000), participants viewed prerecorded videos of others making truthful or deceptive statements. As illustrated in Figure~\ref{fig:stimuli}, we processed the video stimuli through the integrated AI video features of Microsoft Teams to reflect different degrees of AI mediation: in the (1) control condition, the videos were unaltered, corresponding to regular computer-mediated communication; in the (2) weak AI mediation condition, we enabled skin smoothing, lighting adjustment and virtual backgrounds, corresponding to widely used video transformations; in the (3) strong AI mediation condition, we transformed the speaker into a fully-animated character (avatar) to test the effect of strong AI-based video transformation. In addition to a uniform communication setting in which participants rated six videos of the same mediation type in Study 1, participants in Study 2 encountered different types of AI mediation in a mixed environment, more closely resembling real-world interactions. After watching each video, we asked participants how much they trusted the person and whether the person was telling the truth. They also rated their confidence in their judgments and answered follow-up questions about the cues they relied on in their judgments.

Across both studies, we found that AI-mediated video did not affect deception detection accuracy or participants' overall likelihood of suspecting the other person of lying. However, it consistently reduced perceived trustworthiness and lowered participants' confidence in their judgments, particularly in the mixed environment (Study 2). The results suggest that AI-mediated video processing meaningfully affects how people evaluate others in online video communication, even when it does not alter their ability to distinguish truths from lies.

Our findings have implications for platform design and for policy debates, as they show how AI mediation shapes people's evaluations of others in online interactions. Even common tools such as avatars and retouching influence how people form trust, credibility and confidence in online interactions, underscoring the need for greater attention to representational consistency, transparency and the context-sensitive use of mediation features.
\section{Related work}
Our work is motivated by the growing integration of AI-mediated communication tools, such as video retouching and avatars, into everyday video communication \cite{nowakAvatarsComputermediatedCommunication2018,yiRealMeDigital2026}. Prior work in computer-mediated communication \cite{bosEffectsFourComputermediated2002,herringComputermediatedCommunicationInternet2002} has examined how reduced cues shape trust and impression formation, and deception research has documented the limits of people's ability to detect lies \cite{bondAccuracyDeceptionJudgments2006} as well as truth-default rates \cite{levineTruthDefaultTheoryTDT2014}. We draw on this literature and combine it with recent work on AI-mediated communication \cite{hancockAIMediatedCommunicationDefinition2020,hohensteinAIMoralCrumple2020,jakeschAIMediatedCommunicationHow2019}, HCI studies of avatar-mediated communication, and theories of interpersonal judgment, to investigate how AI-mediated video processing affects deception detection, confidence and interpersonal trust.

\subsection{Deception Detection}
Accurately judging whether someone is being honest is essential in mediated interactions \cite{burgoonTrustDeceptionMediated2003}, where people must assess the reliability of information provided by others, such as in hiring conversations, interviews, collaborative online work and educational settings. In many contexts, judgments about honesty shape how people interpret, trust and respond to what is communicated \cite{bondAccuracyDeceptionJudgments2006,levineTruthDefaultTheoryTDT2014}. At the same time, decades of research show that deception detection is difficult \cite{mDeceivingDetectingDeceit1985,vrijDetectingLiesDeceit2008}: people are only slightly better than chance (54\% on average) \cite{bondAccuracyDeceptionJudgments2006,depauloCuesDeception2003} at discerning lies from truth. People also have a strong general tendency to believe what others are saying, known as the "veracity effect" \cite{levineAccuracyDetectingTruths1999,levineTruthDefaultTheoryTDT2014}. 

Previous research has proposed two broad perspectives to explain why deception is complex: On the one hand, cue-based approaches posit that liars reveal themselves through nonverbal cues, so-called "leakage", such as gaze aversion or microexpressions, and that one can discern lies from truth by observing those nonverbal cues \cite{ekmanDetectingDeceptionBody1974,ekmanLyingDeception1997,ekmanNonverbalLeakageClues1969}. However, meta-analyses show that deception cues are inconsistent across studies and are therefore weak and unreliable indicators of deception \cite{depauloCuesDeception2003}. Nonetheless, people still rely on them to form their deception judgments \cite{depauloCuesDeception2003}. In contrast, context-based perspectives, such as Levine's truth-default theory (TDT), argue that people rely primarily on the plausibility and coherence of what is said and default to believing others unless suspicion is actively triggered, which explains the "veracity effect" \cite{levineTruthDefaultTheoryTDT2014}. Cue-based theories and Levine's truth-default theory yield different expectations regarding how the disruption or removal of nonverbal cues in AI-mediated communication might influence deception judgments. If deception cues are central to lie detection, AI systems that modify them could alter truth judgment rate or accuracy. In contrast, if judgments are driven mainly by plausibility and coherence, the AI-based disruption or removal of other elements of the communication may have little to no effect.   

Our work extends prior research on deception detection and AI-mediated communication by introducing AI-mediated videos into established deception stimuli \cite{lloydMiamiUniversityDeception2019}. Despite extensive research on deception detection, no work has examined how everyday AI video tools, such as retouching or avatars, affect our judgments of deception. It also remains unclear how these tools influence our confidence in those judgments and our perceptions of trustworthiness. Particularly, if AI mediation removes or distorts the nonverbal cues that cue-based theories consider central for deception detection, AI-mediated video could make deceptive statements harder to identify. By systematically comparing judgments of videos with varying degrees of AI-mediated content, we provide empirical evidence that AI mediation in video may shape accuracy and interpersonal dynamics in online video communication.

\subsection{Trust in Mediated Communication}

Trust and belief are not the same psychological processes \cite{holtonDecidingTrustComing1994}: While belief is a cognitive judgment about whether a claim is true, trust is a relational, affective stance toward a speaker \cite{holtonDecidingTrustComing1994,hardinTrustTrustworthiness2002}, which is a fundamental precondition for effective human cooperation and interaction \cite{jonesExperienceEvolutionTrust1998,mayerIntegrativeModelOrganizational1995}. High levels of trust enable conflict resolution, problem solving, and fluency in interaction \cite{edmondsonPsychologicalSafetyLearning1999,simonsTaskConflictRelationship2000,zandLeadershipTriadKnowledge1997}, while low trust undermines learning and collaboration \cite{edmondsonPsychologicalSafetyLearning1999,kiffin-petersenTrustIndividualismJob2003}. However, in computer-mediated communication, establishing trust is more difficult than in face-to-face settings \cite{hillOrganizationalContextFacetoface2009,ridingsAntecedentsEffectsTrust2002,roccoTrustBreaksElectronic1998}. Research shows that trust tends to start at a lower baseline in computer-mediated settings and is harder to establish when nonverbal cues are missing \cite{bosEffectsFourComputermediated2002,wilsonAllDueTime2006}. Computer-mediated communication enables people to carefully control how they present themselves, and the related reduction in and manipulation of social cues complicates how trust and credibility are assessed~\cite{ellisonProfilePromiseHonest2013,ertTrustReputationSharing2016,eslamiFirstItThen2016,gillespieMediaTechnologiesEssays2014,hancockDeceptionTechnology2015,hancockTruthLyingOnline2007,herringComputermediatedCommunicationInternet2002}.

The COVID-19 pandemic accelerated the adoption of video communication tools for both casual \cite{hermanFaceExploringCollege2025,javornikWhatLiesFilter2022,vargoDigitalTechnologyUse2021a} and high-stakes settings, including hiring \cite{mccarthyDistressedDistractedCOVID192021,mujtabaFairnessAIDrivenRecruitment2025}, telemedicine \cite{bokoloanthonyjnr.UseTelemedicineVirtual2020,lukasEmergingTelemedicineTools2020,mannCOVID19TransformsHealth2020,sharmaAddressingChallengesAIbased2023}, online exams \cite{EmergenceDeepfakeTechnology2019} and legal proceedings \cite{remusCanRobotsBe2017,vargoDigitalTechnologyUse2021a}. In high-stakes environments, not only does the accuracy of deception judgments matter, but also how credibility and trust are experienced and processed, so understanding how AI-mediated video alters these socio-psychological processes is essential \cite{hancockAIMediatedCommunicationDefinition2020}. 

The integration of artificial intelligence into computer-mediated communication as an additional layer of mediating technologies further increases the sender's control over how they present themselves while potentially complicating judgments on the receiver's side \cite{hancockAIMediatedCommunicationDefinition2020}. Previous work on AI-mediated communication \cite{hancockAIMediatedCommunicationDefinition2020} has shown that, in written contexts, algorithmic modifications complicate how receivers evaluate authenticity and can erode trust and credibility \cite{hancockAIMediatedCommunicationDefinition2020,hohensteinAIMoralCrumple2020,hohensteinAISupportedMessagingInvestigation2018,jakeschAIMediatedCommunicationHow2019,jakeschCoWritingOpinionatedLanguage2023,waltherLetMeCount2005}. This effect is particularly strong in settings where there is a mix of human and AI-generated content, where people start to second-guess others' authenticity, a behavior termed the Replicant Effect \cite{jakeschAIMediatedCommunicationHow2019}. We currently do not understand the extent to which these effects apply to the more dynamic and complex medium of video communication.

Research on AI-mediated video so far has focused on the comparatively extreme cases of generated deceptive video such as deepfakes \cite{baExposingDeceptionUncovering2024,habbalArtificialIntelligenceTrust2024,hancockSocialImpactDeepfakes2021,hwangEffectsDisinformationUsing2021}. Here, AI-generated videos are becoming highly realistic and challenging to distinguish from real footage, so people struggle to tell them apart, which leads to increased uncertainty and reduced trust \cite{hwangEffectsDisinformationUsing2021,popaDeepfakeTechnologyUnveiled2025,EmergenceDeepfakeTechnology2019,twomeyDeepfakeVideosUndermine2023}.
While deepfakes highlight the risks of realistic synthetic video \cite{hancockSocialImpactDeepfakes2021,popaDeepfakeTechnologyUnveiled2025}, much less is known about the impacts of commercial everyday forms of AI mediation in video communication, such as retouching, background replacements or synthetic avatars \cite{boyleEffectsFilteredVideo2000}, offered through widely used video communication platforms like Zoom, Google Meet and Microsoft Teams.

\subsection{Theoretical Mechanisms: Expectancy Violations and Uncertainty Reduction}

We draw on two theoretical perspectives to contextualize a possible decrease in trust in AI-mediated communication. Expectancy Violations Theory (EVT) \cite{burgoonExpectancyViolationsTheory2015} argues that people have internalized assumptions about how others should look and behave in social interaction. When visual or behavioral cues deviate from these internalized expectations in mediated communication---such as when facial features are smoothed, backgrounds are replaced or a speaker is replaced by an animated avatar---people may perceive the interaction as less natural or less aligned with normative social scripts, which can reduce perceived trustworthiness \cite{grimesMentalModelsExpectation2021,hohensteinAIMoralCrumple2020,hongAreYouReady2021,lewSocialScriptsExpectancy2023,rheuWhenChatbotDisappoints2024}. The decrease in trust may be intensified in settings where some people use AI tools and others do not, as differences between mediated and unmediated representations become more salient and thus more likely to violate expectations \cite{burgoonExpectancyViolationsTheory2015,pandaAllTogetherEffectAvatars2022}.

Uncertainty Reduction Theory (URT) offers a complementary perspective on how AI-mediated video may influence confidence in social judgments. Uncertainty Reduction Theory poses that uncertainty in interpersonal encounters is uncomfortable, and that people are motivated to gather information about others to reduce uncertainty and predict others' behavior, attitudes, and intentions \cite{bergerLanguageSocialKnowledge1982,bergerExplorationsInitialInteraction1975}. The motivation to reduce uncertainty is heightened in ambiguous interactions, in which people lack access to the full range of interpersonal cues they typically rely on to minimize uncertainty about another person. In mediated communication, viewers rely on visible signals—such as facial expression, gaze direction, and the timing of responses to reduce ambiguity about a speaker's attitudes or intentions \cite{baxterEngagingTheoriesInterpersonal2008,bergerLanguageSocialKnowledge1982,bergerExplorationsInitialInteraction1975,parksCommunicationNetworksDevelopment1983}. When AI-mediated video alters, smooths or obscures visual cues that could reduce uncertainty in the interaction---e.g., by removing micro-expressions, modifying gaze or reducing facial detail---people might have less information with which to form impressions, increasing uncertainty and lowering the confidence in their own judgments.

\subsection{Avatar-Mediated Communication in HCI}

Avatars are digital representations of users that may be abstract, cartoonish, or human-like. Avatars can be used in online video communication when users prefer privacy, cannot use a camera (e.g., bandwidth or multitasking), or want more control over their visual presentation \cite{nowakAvatarsComputermediatedCommunication2018}. The growing integration of avatars into everyday platforms has motivated HCI researchers to examine how such representations affect presence, expressiveness, and interpersonal evaluation in online meetings \cite{pandaAllTogetherEffectAvatars2022}. A central finding is that high-fidelity avatars are generally more trusted, that is, those avatars that better reflect the real person and their real movements are seen as more trustworthy \cite{pandaAllTogetherEffectAvatars2022}. Ma et al. \cite{maNodsAgreementWebcamDriven2025} find that a critical factor for meeting outcomes is avatar motion fidelity rather than mere visual realism. Webcam-driven head and facial movement supports comfort, emotional clarity, and smooth interaction, while static or synthetic animations make avatars harder to read and increase cognitive load \cite{maNodsAgreementWebcamDriven2025}. In mixed environments particularly, where some people appear via video and others via avatar, the synthetic nature of the avatars becomes more salient \cite{junuzovicSeeNotSee2012,pandaAllTogetherEffectAvatars2022} and people may begin to distrust the avatars \cite{pandaAllTogetherEffectAvatars2022}. Studies have also shown that in professional settings, avatars can interfere with expectations about workplace appropriateness \cite{inkpenMeMyAvatar2011} and that low-fidelity avatars often fall short in conveying facial reactions, gaze and turn-taking cues, leading to lower ratings of professionalism \cite{junuzovicSeeNotSee2012}. 

Research on HCI and avatar-mediated communication highlights that differences in visual fidelity, motion fidelity, and access to nonverbal cues shape how people evaluate others in remote meetings, influencing perceptions of professionalism \cite{inkpenMeMyAvatar2011}, comfort \cite{maNodsAgreementWebcamDriven2025}, and trust \cite{pandaAllTogetherEffectAvatars2022}. However, how such transformations affect judgments of truth or credibility remains an open question. Moreover, the current HCI literature has focused primarily on highly stylized avatar representations \cite{pandaAllTogetherEffectAvatars2022, maNodsAgreementWebcamDriven2025} and has not examined how less sophisticated forms of AI mediation, such as virtual backgrounds, lighting adjustments, or skin retouching, affect people's judgments.
\section{Methods}
To investigate how AI-mediated video, such as retouching, virtual backgrounds or synthetic avatars, affects people's judgments of truthfulness and trust in online video communication, we conducted an experiment simulating an online videoconferencing environment. This section provides details on our experimental design, stimuli, procedure, measurements and recruitment.

\subsection{Hypotheses and Study Design}
Our study design is guided by the larger research question: "Does AI-mediated video communication disrupt or interfere with deception judgments and interpersonal trust in online communication?" Based on prior research on deception detection and AI-mediated and avatar-mediated communication, we formulated five hypotheses about how mediating AI systems might affect people's ability to detect lies, confidence in their judgments and trust in others. For each hypothesis, we outline the relevant theoretical mechanisms that motivate the predictions and their direction.

Trust is a central component of interpersonal evaluation in mediated communication \cite{hillOrganizationalContextFacetoface2009,ridingsAntecedentsEffectsTrust2002,roccoTrustBreaksElectronic1998}. Here, prior CMC research shows that the reductions or distortions of interpersonal cues can decrease perceived trustworthiness \cite{bosEffectsFourComputermediated2002,wilsonAllDueTime2006}. Expectancy Violations Theory (EVT) \cite{burgoonExpectancyViolationsTheory2015} suggests that visual changes such as avatars or retouching filters may reduce trust when they deviate from what people anticipate as "normal" in a video-call setting. At the same time, research on avatar-mediated communication shows that some mediated representations—particularly those perceived as appropriate or expressive—can be evaluated positively \cite{inkpenMeMyAvatar2011}. As prior work indicates that trust could increase or decrease depending on the nature and quality of AI mediation, we formulate the following non-directional hypothesis:

\begin{enumerate}[label=\textbf{H\arabic*}]
    \item \textbf{Interpersonal Trust:} AI-mediated video affects the perceived trustworthiness of the speaker.
\end{enumerate}

Truth judgment rate refer to how often viewers judge statements as true. They are a central outcome in deception detection research \cite{bondAccuracyDeceptionJudgments2006}. Cue-based perspectives propose that the cognitive effort of lying produces micro-expressions that the receiver can observe to detect deception \cite{ekmanLyingDeception1997}. If AI mediation reduces access to nonverbal cues, for example, by replacing the speaker with a synthetic avatar, viewers may judge statements as more truthful because they fail to detect cues of deception. In contrast, Levine's truth-default theory argues that people generally default to judging statements as true unless suspicion is triggered \cite{levineTruthDefaultTheoryTDT2014}. AI mediation could increase uncertainty or suspicion and result in breaking out of that truth-default state, which would reduce truth judgment rate; that is, people would believe the AI-mediated speaker less often, so we hypothesize that:

\begin{enumerate}[label=\textbf{H\arabic*}, resume]
    \item \textbf{Truth judgment rate:} AI-mediated video affects the rate at which participants judge what is said as true.
\end{enumerate}

In deception detection research, humans achieve only slightly better-than-chance accuracy (54\%) at correctly classifying veracity judgments \cite{bondAccuracyDeceptionJudgments2006}. Similar to truth judgment rate (H2), cue-based theories might predict that removing or disrupting visual cues used to detect deception could impair people's ability to make accurate truth-lie judgments, thereby reducing accuracy. In contrast, Levine's truth-default theory predicts that accuracy could improve when people rely less on nonverbal cues and more on message content \cite{levineTruthDefaultTheoryTDT2014}. Reducing visible cues under AI mediation could strengthen the intuition to shift to content-based information rather than visual cues, thereby increasing accuracy. Furthermore, meta-analyses show that accuracy often remains unchanged regardless of viewing conditions \cite{bondAccuracyDeceptionJudgments2006,depauloCuesDeception2003}, which would predict a flat performance in accuracy across all levels of AI mediation. We hypothesize:

\begin{enumerate}[label=\textbf{H\arabic*}, resume]
    \item \textbf{Judgment accuracy:} AI-mediated video affects the rate at which participants judge truths as truths (truth accuracy, true positives) and lies as lies (lie accuracy, true negatives).
\end{enumerate}

Confidence in judgments is a subjective metric that captures how sure people feel about their judgments \cite{depauloAccuracyConfidenceCorrelationDetection1997}. Uncertainty Reduction Theory (URT) \cite{bergerLanguageSocialKnowledge1982,bergerExplorationsInitialInteraction1975} suggests that when familiar interpersonal cues (e.g., facial expressions) are reduced or altered, uncertainty increases, which may lower people’s confidence in their own judgments in AI-mediated interactions. At the same time, prior avatar and mediated-communication studies suggest that when representations are appropriate or easy to interpret, people may feel more certain in their evaluations \cite{pandaAllTogetherEffectAvatars2022,maNodsAgreementWebcamDriven2025}. As prior lines of work predict both decreases and increases in confidence, we propose the following non-directional hypothesis:

\begin{enumerate}[label=\textbf{H\arabic*}, resume]
    \item \textbf{Judgment confidence:} AI-mediated video affects participants' confidence in their deception judgments.
\end{enumerate}

In real-world interactions, people may often encounter a mix of original video communications interspersed with retouch effects, virtual backgrounds and avatars. Such mixed environments may increase the salience of the AI mediation and may affect how participants react to the use of AI \cite{jakeschAIMediatedCommunicationHow2019}. Prior HCI and avatar-mediation work shows that differences in environment influence how people evaluate one another \cite{pandaAllTogetherEffectAvatars2022}. From a theoretical perspective, Expectancy Violations Theory (EVT) \cite{burgoonExpectancyViolationsTheory2015} suggests that a mixed environment where people communicate with different levels of AI mediation alongside original video representations may reinforce both expectation violations and a sense of uncertainty. We hypothesize: 

\begin{enumerate}[label=\textbf{H\arabic*}, resume]
    \item \textbf{Interaction with type of environment:} The impact of AI-mediated video on accuracy, trust and confidence is stronger in settings where people see a mix of different types of AI mediation compared to settings where everyone uses the same AI tools.
\end{enumerate}

To test the above hypotheses empirically, we designed a study consisting of two experiments: a between-subjects experiment (Study 1) to test H1 to H4 in an environment where participants encounter a single type of AI mediation only; and a within-subjects experiment (Study 2), where participants encounter different types of AI mediation to test H5 in addition to H1 to H4. We preregistered the hypotheses together with the study design and analysis plan before data collection \footnote{\href{https://aspredicted.org/download_pdf.php?a=dDhwc09ZMWlvMEFvU0p4bG1ZcjZpZz09&t=aEpOYzhadWlOY21JK01XdGc1SkZsUT09}{AI Video Filters and Deception Judgments – Study 1 \& 2 Preregistration (\#239571), submitted 2025-07-23 on AsPredicted}.}.

\subsection{Stimuli and Experimental Treatments}
We structured the experiment as two complementary studies conducted concurrently. In the experiments, participants evaluated six videos in a videoconferencing platform setting. We processed the videos to different levels of AI-based transformations, with video subjects telling true or fabricated stories about another person. We asked participants to judge whether video subjects were telling the truth or lying. 

We considered several video deception datasets for studying deception across different lie stakes. The Miami University Deception Detection Database (MU3D) \cite{lloydMiamiUniversityDeception2019} provides truthful and deceptive videos. The Bag-of-Lies dataset \cite{guptaBagOfLiesMultimodalDataset2019a} includes multimodal signals such as video, audio, and biometrics in low to medium-stakes laboratory settings. DOLOS \cite{guoAudioVisualDeceptionDetection2023} presents medium-stakes deception from incentivized game-show interactions with richly annotated audiovisual data. Other high-stakes datasets include courtroom trial recordings \cite{perez-rosasDeceptionDetectionUsing2015} and political deception videos \cite{walkerMergingAIIncidents2024}, where consequences add complexity. 

For our present work, we used the Miami University Deception Detection Database \cite{lloydMiamiUniversityDeception2019}, which contains webcam-recorded videos of 80 subjects, equally divided by race (black/white) and gender (male/female). Each subject is featured in four videos, in which they either make a truthful or a deceptive statement about their social relationships, under a positive or negative valence. The dataset captures unscripted, conversational speech with direct camera eye contact and natural behavior, closely mimicking the dynamics of online video communication platforms and aligning well with the purpose of the current study. As the valence dimension is irrelevant to the current study and would have introduced unnecessary variation, we focused on positive-valence videos only, yielding a 160-video base set comprising 80 lies and 80 truths.

We embedded these videos into a video communication platform, Microsoft Teams, and further processed them to reflect different levels of AI mediation (see Figure \ref{fig:stimuli}) in addition to the control condition:

\begin{enumerate}[label=\textbf{T\arabic*}]
    \item \textbf{Control condition:} In the control condition, participants saw the original, unaltered recording embedded in a Microsoft Teams interface. 
    \item \textbf{Weak AI mediation condition:} In the weak AI mediation treatment, the videos were preprocessed with the skin smoothing, lighting adjustments and virtual backgrounds features offered by Microsoft Teams. 
    \item \textbf{Strong AI mediation condition:} In the strong AI mediation treatment, we further processed the videos with Microsoft Teams' avatar feature that replaced the person in the video with a fully synthetic representation of the speaker. For the treatment, 80 digital avatars were manually designed in the Microsoft Avatars App by the first author and a research assistant to resemble the speaker in the original video. 
\end{enumerate}

By using the integrated video-processing features of a video communication tool commonly used in professional settings, we can study comparatively common and realistic stimuli. In contrast to previous work studying more extreme forms of AI-mediated video communication, such as deepfakes \cite{hancockSocialImpactDeepfakes2021,twomeyDeepfakeVideosUndermine2023}, our treatments provide insights into a communication setting encountered daily by millions of people. The calibration of our treatments was further informed by in-person testing and an initial pilot study with a subset of 16 videos from 8 video subjects and N = 100 participants.

\subsection{Procedure and Measurements}
Before beginning the study, all participants provided informed consent and read brief instructions explaining the main task. They also completed an attention-check question to confirm their understanding of the task before proceeding to the main task. The main task consisted of evaluating six short, prerecorded video clips (approximately 40 seconds long) \cite{lloydMiamiUniversityDeception2019}, preprocessed according to the treatment conditions described above. Videos were balanced for veracity, ensuring that each participant viewed exactly three truthful and three deceptive statements. Additionally, we balanced video subjects across gender and race. Although we had multiple videos per speaker, we ensured that each participant saw each speaker at most once. Participants could watch each video only once using standard playback controls, except for the replay option. We limited the number of videos to six per participant to balance sufficient exposure to each condition while minimizing participant fatigue.

After each video, participants answered three questions:
\begin{enumerate}[label=\textbf{O\arabic*}]
    \item \textbf{Veracity Judgment:} "Do you think this person is lying or telling the truth?" (Binary choice: Lie / Truth)
    \item \textbf{Judgment Confidence:} "How confident are you in your judgment?" (5-point Likert scale: Not at all confident to Extremely confident)
    \item \textbf{Trustworthiness:} "How trustworthy does the person in the video seem?" (5-point Likert scale: Not at all trustworthy to Extremely trustworthy)
\end{enumerate}
We adjusted the scale directions across participants to mitigate order bias. In addition to the three main outcome variables of judgment, judgment confidence and trustworthiness, participants also answered an open-ended exploratory question for the last video only, where they explained their judgment ("Why do you think this person is lying or telling the truth?") and completed a multiple-choice question indicating what specific cues influenced their judgment ("Which of the following most influenced your judgment of whether the person was telling the truth?"). We allowed participants to select up to three cues from a set of eight cues we assembled based on a review of categories of cues reported in prior deception research: visual nonverbal cues, vocal/paraverbal cues, content-based cues and global demeanor \cite{levineScientificEvidenceCue,depauloCuesDeception2003}. 

Finally, participants estimated their own judgment success ("Out of the six videos you watched, how many do you believe you correctly assessed?"). As a manipulation check, participants also indicated how many of the six videos they believed featured an avatar. They answered an open-ended exploratory question ("Did you notice anything unusual or artificial about the videos?"). The study concluded with demographic and exploratory questions about participants' use of online video communication platforms and their experience with AI-based video features.
After submitting demographic information, participants received a detailed debriefing statement explaining the study's purpose. 

\subsection{Analysis Approach}
We analyzed trust (H1), truth judgment rate (H2), accuracy (H3) and judgment confidence (H4) using separate linear mixed-effects models for each study, with AI mediation level as a fixed effect and random intercepts for participants to account for repeated measurements. To test the effect of the environment on trust, truth judgment rate, accuracy, and confidence (H5), we fitted a combined mixed-effects model across both studies with a fixed-effect interaction between AI mediation level (weak or strong) and environment type (homogeneous or mixed). As a further robustness check, we also estimated an extended version of these models that included demographic and experience-related covariates: age, gender, education, race, English proficiency, prior experience with video tools, prior experience with AI tools, frequency of AI interaction, and general trust in AI. We report descriptive statistics of means and standard deviations alongside model estimates ($\beta$ coefficients, 95\% CIs, p-values). Hypotheses are evaluated based on the model results.

\subsection{Recruitment}
In Study 1 (between-subjects design), participants were randomly assigned to one of the treatment conditions and evaluated six videos of the same type. The between-subjects design isolates the impact of each manipulation, allowing for robust conceptual comparisons. In Study 2 (within-subjects design), participants viewed two videos per condition in random order, reflecting real-world variability in mediated communication and allowing us to capture how participants reacted to AI features in settings where some, but not all, people use them. To allow for valid comparisons across studies, participants for Study 1 and Study 2 were recruited concurrently from the same sample and randomly assigned to one of the two study designs.

A total of 2,000 participants were recruited through Prolific \cite{palanProlificacASubjectPool2018}. Eligibility criteria required that participants be 18 years or older and be resident in the United States. We determined the sample size based on a bootstrapped power analysis conducted before data collection. 
Note that our sample differs from the preregistration, as we initially planned to recruit only 1,000 participants. The initial power analysis, based on pilot study data, estimated the required sample size to achieve approximately 80\% power to detect small changes (d = .2) in trust and confidence.  
After collecting an initial 1,000 answers, we realized that relevant effect sizes in accuracy (3-5\%, corresponding to a Cohen's d = .06 to .1) were substantially smaller than estimated, and the initially planned sample would leave us with variations in accuracy that were difficult to interpret. Increasing the sample size to N = 2,000 enabled us to detect larger accuracy differences (d = .1) with approximately 70\% power. The larger sample should also have improved the robustness and interpretability of our overall study. As no changes in accuracy were detected with the increased sample, we see no risk of false positives.

We compensated participants \$2 for their participation, which, on average, took about 10 minutes, corresponding to a \$12 hourly rate. To encourage attentive responding and to raise the stakes of the scenario, we offered participants a \$2 bonus if they classified at least 5 videos correctly, doubling their base payment. 289 participants received the bonus payment. 
Participants ranged in age from 21 to 77 (M = 41, SD = 13.7, Median = 37). Male participants represented 61.3\% and female participants represented 38.7\% of the sample. 65.8\% of the sample self-identified as White or Caucasian; 22.5\% as Black or African American; 4.5\% as Asian; 4.5\% as Latino or Hispanic; and 2.7\% as Indigenous, Middle Eastern, North African, or mixed race. Most participants were highly educated, with 42.3\% holding a bachelor's degree and 23.4\% a master's degree. The majority were native English video subjects (90.1\%), with the remainder reporting advanced or intermediate English proficiency.

\section{Results}
In this section, we present the empirical results from the experiments and analyze how AI-mediated video processing influenced interpersonal trust, judgment accuracy and confidence. The results reveal that AI mediation affects perceptions of trust and the confidence with which people make judgments, but has limited effects on actual judgment accuracy or on the tendency to believe others.

\begin{figure*}
  \begin{center}
    \includegraphics[width=.65\textwidth, trim=0.2cm 0.2cm 0.2cm 0, clip]{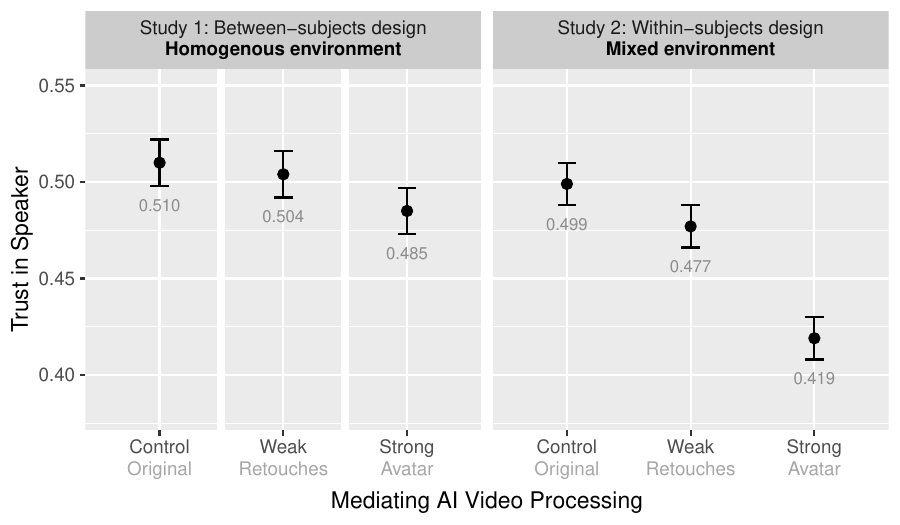}
  \end{center}
\caption{\textbf{AI mediation affects interpersonal trust, particularly in mixed environments.} \textit{Average interpersonal trust by video mediation type and study design with 95\% confidence intervals; N = 2,000 ratings per data point.} Participants in the left panel (Study 1) rated six videos of the same mediation type, whereas participants in the right panel (Study 2) watched two videos of each mediation type in random order.
}
\Description{AI mediation affects interpersonal trust, particularly in mixed environments. Average interpersonal trust by video mediation type and study design with 95\% confidence intervals; N = 2,000 ratings per data point. Participants in the left panel (Study 1) rated six videos of the same mediation type, whereas participants in the right panel (Study 2) watched two videos of each mediation type in random order.}
\label{fig:trust}
\end{figure*}

\textbf{Interpersonal trust (H1):}Figure \ref{fig:trust} shows participants' trust in the person in the video across different types of AI mediation. The left panel shows trust ratings in homogeneous environments (Study 1), where participants encountered only one type of mediation, with the different types of mediation shown on the x-axis. In the control condition on the far left, where participants saw only original videos without AI mediation, the video subjects received an average trust rating of 0.51 (SD = 0.265), corresponding to "moderately trustworthy". Subjects using retouch effects received similar trust ratings (M = 0.504, SD = 0.263), while video subjects using avatar filters received slightly lower trust ratings (M = 0.485, SD = 0.257). 
We fitted a linear mixed model to predict reported trust by mediation type in Study 1 with a per-participant random fixed effect to account for the repeated measures design. The effect of avatar-based mediation on trust is statistically significant and negative ($\beta$ = -0.03, 95\% CI [-0.05, -0.002], t(5797) = -2.15, p = .032). In contrast, the effect of the weak AI mediation condition is statistically non-significant ($\beta$ = -0.006, 95\% CI [-0.03, 0.02], t(5797) = -0.52, p = .604). We provide further details on the model in Table \ref{table-model1} in the Appendix.

In Study 2, where participants encountered different types of AI mediation in a mixed environment (right panel in Figure~\ref{fig:trust}), the effect of AI mediation on trust was more substantial. Video subjects in the original video, shown in the left column, received an average trust rating of 0.499 (SD = 0.264), similar to the trust ratings in the Study 1 control group. Video subjects who used retouches with virtual backgrounds received slightly lower trust ratings (M = 0.477, SD = 0.259), whereas video subjects who used avatars as the stronger AI mediation received substantially lower trust ratings (M = 0.419, SD = 0.264). We fitted a linear mixed model with a per-participant random fixed effect to predict reported trust by mediation type in Study 2. Compared to the control condition, the effect of AI mediation on perceived trust of the speaker in the mixed environment is statistically significant and negative for the retouch condition (M = .477, $\beta$ = -0.02, 95\% CI [-0.04, -0.006], t(6235) = -2.81, p = .005) and for avatars (M = 0.419, $\beta$ = -0.08, 95\% CI [-0.09, -0.07], t(6235) = -10.51, p < .001). We provide further details on the model in Table \ref{table-model2} in the Appendix.

Next, we analyzed the extent to which the effects of different types of AI mediation differed across environments by comparing results from Study 1 and Study 2 (H5).
We fitted a linear mixed model predicting reported trust, with mediation type and environment type as predictors, across Study 1 and Study 2, including a per-participant random fixed effect to account for the repeated-measures design. 
    
    
The interaction term for the avatar condition was statistically significant and negative, indicating that the trust penalty for avatars was substantially larger in mixed environments than in homogeneous ones ($\beta$ = -0.05, 95\% CI [-0.08, -0.03], t(12034) = -4.05, p <
.001). We found no reliable interaction for retouch videos ($\beta$ = -0.02, 95\% CI [-0.04, 0.01], t(12034) = -1.13, p = .257). The main effect of the environment was non-significant ($\beta$ = -0.01, 95\% CI [-0.03, 0.008], t(12034) = -1.14, p = .253), showing that baseline trust in the control condition did not differ between environments. 
We provide further details on the model in Table \ref{table-model3} in the Appendix.

Overall, the results support the hypothesis that AI-mediated video influences perceived trustworthiness (H1). Considering the effect of the environment (H5), trust in avatar video subjects in mixed environments was significantly lower than in the homogeneous setting. 

\begin{figure*}
  \begin{center}
    \includegraphics[width=.65\textwidth, trim=0.2cm 0.2cm 0.2cm 0, clip]{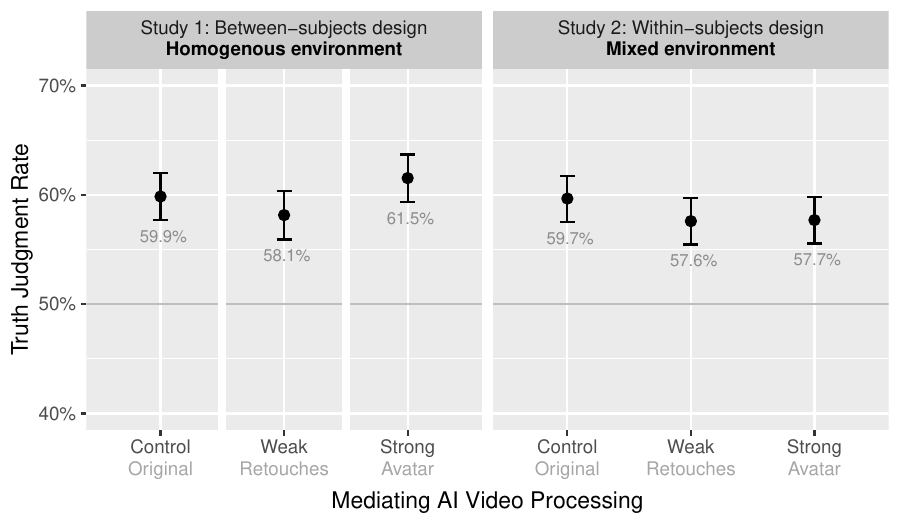}
  \end{center}
\caption{\textbf{Truth judgment rates were unaffected by mediating AI mediation} \textit{Percentage of times participants thought the person in the video told the truth, with 95\% Wilson confidence intervals; N = 2,000 judgments per data point.} Across both studies, participants exhibit a bias towards truth judgment rate (57.6 to 61.5\%) independent of AI mediation, while the stimuli dataset contained 50\% truths and 50\% lies.
}
\Description{Truth judgment rates were unaffected by mediating AI mediation. Percentage of times participants thought the person in the video told the truth, with 95\% Wilson confidence intervals; N = 2,000 judgments per data point. Across both studies, participants exhibit a bias towards judging statements as true (57.6\% to 61.5\%) independent of AI mediation, while the stimuli dataset contained 50\% truths and 50\% lies.}
\label{fig:truth_rate}
\end{figure*}

\textbf{Truth judgment rate (H2):} Although we observed a reduction in trust through AI mediation, we did not observe changes in how often participants thought the person in the video was lying. Figure \ref{fig:truth_rate} shows participants' truth judgment rate, that is, how often participants indicated that they thought the person in the video was telling the truth. In the control condition, in which participants were shown the original video, they believed video subjects were telling the truth about 60\% of the time (59.9\% in Study 1 and 59.7\% in Study 2). Note that this rate is significantly higher than the ground truth frequency shown in grey, aligning with other studies showing that people are truth-biased \cite{levineTruthDefaultTheoryTDT2014,levineAccuracyDetectingTruths1999}. However, we observed no significant difference in truth judgment rates when video subjects used retouches (58.1\% and 57.6\%) or avatar filters (61.5\% and 57.7\%) across both studies. Given our overall sample size, we would have expected to detect a change of about 3-5\% most of the time.

We fitted linear mixed models to predict reported truth judgment rate by mediation type in Study 1 and in Study 2 with a per-participant random fixed effect. In neither study was the effect of avatar-based mediation on truth judgment rate statistically significant. The effect of weak AI mediation (retouch) on truth judgment rate is statistically non-significant in Study 1 ($\beta$ = -0.02, 95\% CI [-0.05, 0.01], t(5797) = -1.09, p = .278) and Study 2 ($\beta$ = -0.02, 95\% CI [-0.05, 0.009], t(6235) = -1.35, p = .176), as is the effect in the avatar condition for Study 1 ($\beta$ = 0.02, 95\% CI [-0.01, 0.05], t(5797) = 1.07, p = .284) and Study 2 ($\beta$ = -0.02, 95\% CI [-0.05, 0.01], t(6235) = -1.29, p = .19). Furthermore, we fitted a linear mixed model to predict the reported truth judgment based on mediation type and environment type across Study 1 and Study 2 with a per-participant random fixed effect. None of the interaction terms were statistically significant, indicating no effects on the type of environment (H5). We provide further details on the models in Table \ref{table-model1}, \ref{table-model2} and \ref{table-model3} in the Appendix.

We observed no differences in truth judgment rates across mediation types and environments. The results do not support H2 but are consistent with Levine's truth-default theory, which posits that people have a consistent truth bias and default to believe others unless clear suspicion is triggered \cite{levineTruthDefaultTheoryTDT2014}. 

\begin{figure*}
  \begin{center}
    \includegraphics[width=.84\textwidth, trim=0.2cm 0.2cm 0.2cm 0, clip]{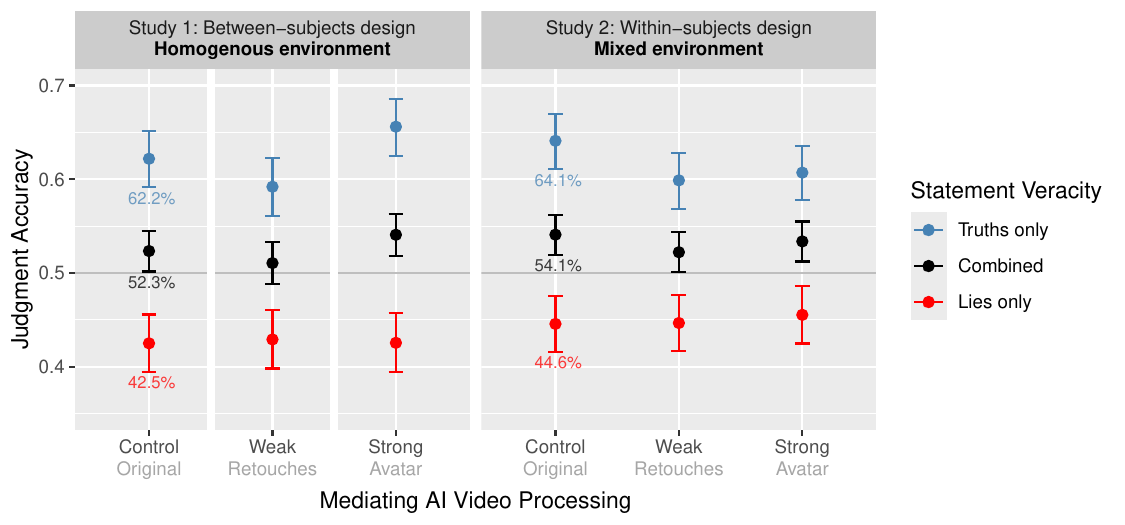}
  \end{center}
\caption{\textbf{Judgment accuracy by AI mediation type and statement veracity.} \textit{Percentage of times participants correctly identified a truth as a truth or a lie as a lie, with 95\% Wilson confidence intervals; N = 1,000-2,000 judgments per data point.} Participants identified about 60-65\% of truths correctly, but only about 42-45\% of lies. Except for truths in the avatar treatment in Study 1, AI mediation did not affect judgment accuracy.}
\Description{Judgment accuracy by AI mediation type and statement veracity. Percentage of times participants correctly identified a truth as a truth or a lie as a lie, with 95\% Wilson confidence intervals; N = 1,000-2,000 judgments per data point. Participants identified about 60-65\% of truths correctly, but only about 42-45\% of lies. Except for truths in the avatar treatment in Study 1, AI mediation did not affect judgment accuracy.}
\label{fig:accuracy}
\end{figure*}

\textbf{Judgment accuracy (H3):} Figure \ref{fig:accuracy} shows participants' judgment accuracy across conditions, that is, how often they rated truths as truths and lies as lies. The black data points at the center of the graph show the overall accuracy on the combined set of stimuli, half of which contained truths and half of which contained lies. The grey reference line shows the baseline accuracy that participants would have achieved by random responses (50\%). In line with findings in related work, participants were slightly better than random at telling truths from lies, with 52.3\% accuracy in the Study 1 control group and 54.1\% in the control group of Study 2. This rate did not change significantly across videos with retouching (51.1\% and 52.2\%) or avatar (54.1\% and 53.4\%) in Study 1 and Study 2.

We fitted two linear mixed models to predict judgment accuracy by mediation type in Study 1 and Study 2, with a per-participant random fixed effect. Participants achieved higher accuracy on videos in which subjects told the truth (shown in blue), with 62.2\% in the control condition in Study 1 and 64.1\% in the control condition in Study 2. While this level of accuracy is significantly higher ($\beta$ = 0.20, 95\% CI [0.17, 0.23], p < .001) than the accuracies participants achieved in videos with lies (42.5\% and 44.5\% respectively), this difference largely reflects the general truth bias in participants' judgments observed above. In Study 2, we observed a non-significant reduction in accuracy in the retouch condition (59.9\%, $\beta$ = -0.02, 95\% CI [-0.05, 0.01], t(6235) = -1.22, p = .224). 

We tested the mediating effect of the environment type (H5) by fitting a linear mixed-effects model predicting accuracy from mediation type, study environment and their interaction. The model showed no statistically significant effects of either retouch ($\beta$ = –0.007, 95\% CI [–0.08, 0.06], p = .844) or avatar mediation ($\beta$ = 0.04, 95\% CI [–0.03, 0.11], p = .238) relative to the control condition. The interaction terms assessing whether mediation effects differed between homogeneous and mixed environments were also non-significant for both retouch ($\beta$ = –0.006, 95\% CI [–0.05, 0.04], p = .791) and avatar conditions ($\beta$ = –0.02, 95\% CI [–0.07, 0.02], p = .270). We provide the full model details in Table \ref{table-model3} in the Appendix.

Participants achieved 51–54.5\% overall accuracy, corresponding to just 1–4.5\% above chance. Consequently, only relatively large AI mediation effects that meaningfully improved or reduced this limited margin of accuracy would be detectable. Although our studies were adequately powered to detect changes of 3–5\%, they were not sufficiently sensitive to reliably capture smaller effects of 1–2\%. Our results show no meaningful differences in accuracy across mediation types (H3) or environments (H5), aligning with prior meta-analysis work on deception detection research showing that human deception detection accuracy remains stable across viewing conditions and is only slightly above chance \cite{bondAccuracyDeceptionJudgments2006,depauloCuesDeception2003}.

\begin{figure*}
  \begin{center}
    \includegraphics[width=.65\textwidth, trim=0.2cm 0.2cm 0.2cm 0, clip]{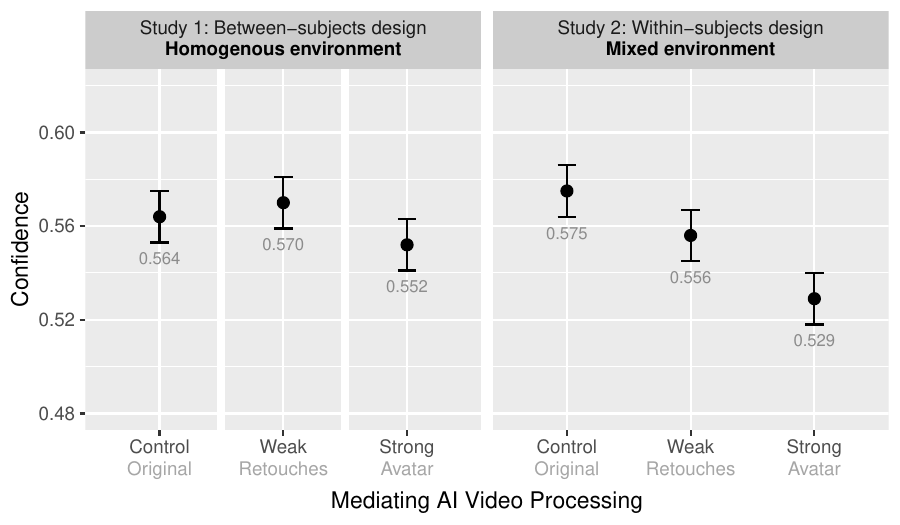}
  \end{center}
\caption{\textbf{AI mediation affects judgment confidence, but only in mixed environments.} \textit{Average reported confidence in judgment by video mediation type and study design with 95\% confidence intervals; N = 2,000 ratings per data point. } Participants encountering different types of AI-mediated content and original content in Study 2 (right panel) were less confident in their judgments, particularly for strong AI-mediated content. 
}
\Description{AI mediation affects judgment confidence, but only in mixed environments. Average reported confidence in judgment by video mediation type and study design with 95\% confidence intervals; N = 2,000 ratings per data point. Participants encountering different types of AI-mediated and original content in Study 2 (right panel) were less confident in their judgments, particularly for strong AI-mediated content.}
\label{fig:confidence}
\end{figure*}

\textbf{Judgment confidence (H4):} We fitted a linear mixed model to predict the reported judgment confidence based on mediation type in Study 1 and Study 2 with a per-participant random fixed effect to account for the repeated measures design. While AI mediation did not affect participants' judgment accuracy, it did affect their confidence in their judgments, as shown in Figure \ref{fig:confidence}. In the control conditions, participants reported a mean confidence of 0.56 (SD = 0.258) in Study 1 and 0.57 (SD = 0.253) in Study 2, which falls between moderately confident (0.5) and very confident (0.75) on the Likert scale. In Study 1, where participants encountered only one type of mediation, their confidence did not change significantly when they encountered videos with retouching (M = 0.57, SD = 0.255, $\beta$ = 0.006, 95\% CI [-0.02, 0.03], t(5797) = 0.44, p = .661) or avatar filters (M = 0.55, SD = 0.248, $\beta$ = -0.01, 95\% CI [-0.04, 0.02], t(5797) = -0.86, p = .391). In Study 2, however, where participants encountered a mix of different mediation types, they were significantly less confident in their judgments when evaluating videos with retouches (M = 0.556, SD = 0.249) and avatar filters (M = 0.529, SD = 0.258). Linear mixed-effects models confirm a significant negative effect compared to the control condition for retouch (M = 0.556, $\beta$ = -0.02, 95\% CI [-0.03, -0.006], t(6235) = -2.94, p = .003) and avatar filters (M = 0.529, $\beta$ = -0.05, 95\% CI [-0.06, -0.03], t(6235) = -7.10, p < .001). Details on the models are provided in Table \ref{table-model1} and Table \ref{table-model2} in the Appendix.

To test whether the effect of AI mediation on confidence differs between homogeneous and mixed environments (H5), we fitted a linear mixed-effects model with an interaction between level of AI mediation and environment, including a random intercept for participants (see Table \ref{table-model3}). While the interaction for retouches was not statistically significant ($\beta$ = -0.03, 95\% CI [-0.05, 0.0043], t(12034) = -1.67, p = .095), the model shows a statistically significant interaction in the avatar condition ($\beta$ = -0.03, 95\% CI [-0.06, -0.004], t(12034) = -2.23, p = .026), suggesting that confidence drops more sharply for avatar-mediated videos in mixed environments than in homogeneous ones, aligning with findings from prior work on HCI and avatar-mediated communication research \cite{pandaAllTogetherEffectAvatars2022}.

Judgment confidence decreased for AI-mediated video subjects, particularly for avatars and in mixed environments, consistent with H4 and H5. While confidence remained stable in homogeneous settings (Study 1), it declined in mixed environments in which participants had to compare differently mediated videos side by side, aligning with prior work on avatar mediation \cite{pandaAllTogetherEffectAvatars2022} and Uncertainty Reduction Theory. 

\begin{figure*}
  \begin{center}
    \includegraphics[width=.8\textwidth, trim=0.2cm 0.2cm 0.2cm 0, clip]{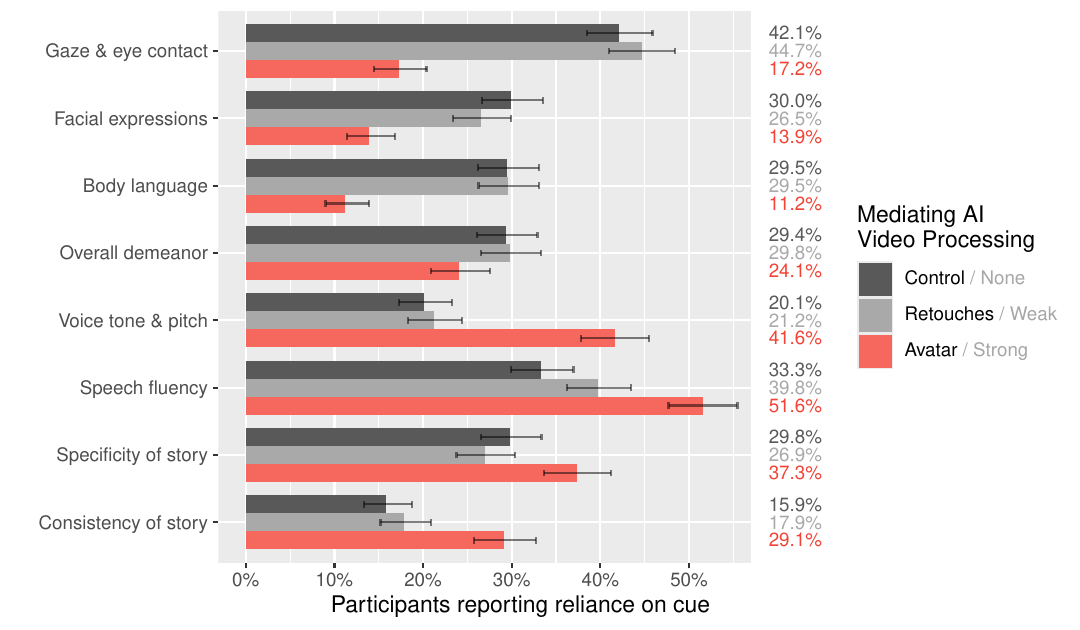}
  \end{center}
\caption{\textbf{Participants relied more on content-based cues and less on expressions and body language when evaluating AI-mediated content.} \textit{Percentage of participants indicating reliance on different cues based on mediation type, with 95\% Wilson confidence intervals; N = 2,000.} Participants relied substantially on gaze, expressions and body language (top three rows) when evaluating the original and retouched videos, cues that were lost under strong AI mediation. Instead, participants relied more on voice, fluency, and content consistency (bottom four rows) when evaluating avatar videos.
}
\Description{Participants relied more on content-based cues and less on expressions and body language when evaluating AI-mediated content. Percentage of participants indicating reliance on different cues based on mediation type, with 95\% Wilson confidence intervals; N = 2,000. Participants relied substantially on gaze, expressions and body language (top three rows) when evaluating the original and retouched videos, cues that were lost under strong AI mediation. Instead, participants relied more on voice, fluency, and content consistency (the bottom four rows) when evaluating videos featuring avatars.}
\label{fig:cues}
\end{figure*}

Figure \ref{fig:cues} summarizes the answers participants gave when asked which cues or elements most influenced their judgments in the last video. We coded each cue option as a binary indicator (selected = 1, not selected = 0) and report the percentage of participants selecting each cue along with Wilson 95\% confidence intervals. In a multiple-choice question, participants could select up to three cues, such as gaze and eye contact, body language, or speech fluency, shown on the y-axis on the left of Figure~\ref{fig:cues}. The x-axis shows how often participants selected a cue, depending on the AI mediation type of the relevant video. Control and retouch videos are shown in dark and light grey, respectively, and avatar videos in red. We gathered the set of cues available from three major categories of deception research to cover the most reported cues according to prior work: visual nonverbal cues, vocal or paraverbal cues, content-based cues and global demeanor \cite{levineScientificEvidenceCue,depauloCuesDeception2003}.

In the control condition, participants substantially relied on gaze and eye contact (42.1\%), facial expressions (30\%), and body language (29.5\%) to make their judgments. The retouch condition (weak AI mediation) closely resembles the control condition in the use of nonverbal cues such as gaze and eye contact (44.7\%), facial expressions (26.5\%), and body language (29.5\%). By contrast, when the speaker used an avatar instead, participants reported relying less on these nonverbal cues (17.2\%, 13.9\%, and 11.2\%, respectively). Instead, participants shifted their attention from visual cues to audio and content-based cues, relying on voice tone and pitch (41.6\%), speech fluency (51.6\%), story specificity (37.3\%), and consistency (29.1\%) more than participants in the control group (20.1\%, 33.3\%, 29.8\%, and 15.9\%, respectively). Overall, these results show that while AI mediation did not affect participants' judgment accuracy, it did change how they arrived at their judgments and how confident they felt about them, providing mechanisms that help explain why participants felt less confident in their judgments and trusted the speaker less in strongly AI-mediated videos.

\subsection{Manipulation Checks}
As a manipulation check, we asked participants to indicate how many of the six videos they watched featured a fully-animated avatar after the main task. In Study 2 (mixed environment), participants rated exactly two avatar videos and, on average, reported a similar number (M = 2.22, SD = 0.58). In Study 1 (homogeneous environment), participants in the avatar condition (strong AI mediation) correctly recognized that all or most of the videos featured avatars (M = 5.78, SD = 0.86). 
In the open-ended question ("Did you notice anything unusual or artificial about the videos?"), participants also frequently commented on the avatar videos in the mixed environment (Study 2), describing them as "not a real person," "weird," "off," "unnatural," or "hard to read." Several participants noted that the avatar "looked artificial" or that the speaker disappeared to be "behind an avatar": "I immediately don't trust people who use avatars. They are generally 'social snipers' who hide behind anonymity so they don't have to be responsible for their actions" (Participant N474). Comments on the virtual background or visual filters from the retouching condition were less frequent. They typically referred to mild visual artifacts, such as "blurred background," "border around the person," or "the lighting seemed edited".

Overall, our two manipulation checks suggest that strong AI mediation was salient to participants, particularly in the mixed environment of Study 2. The qualitative responses indicate that participants noticed strong AI-mediated videos and occasionally noted minor artifacts in weak mediation conditions. 

\subsection{Robustness Checks}
To increase the robustness of our results, we fitted an extended linear mixed-effects model across both studies with covariates including age, gender, race, education, English proficiency, subjective trust in AI systems, and prior experience with online video communication platforms, video filters and AI mediation, in addition to the independent variables of mediation type and study environment (see Appendix Table \ref{table-model4}).

All significant effects reported in the previous studies remain significant in the extended model, after including the covariates: the model predicts that the negative effect of avatars on trust remains significant ($\beta$ = –0.027, p < .05), particularly in the mixed environment ($\beta$ = –0.080, p < .001). A somewhat smaller but statistically significant trust reduction was also predicted for retouched videos in mixed environments ($\beta$ = –0.021, p < .01). Similarly, a significant decrease in confidence was predicted for retouched videos in mixed environments ($\beta$ = –0.019, p < .01), with an even larger reduction for avatars in mixed environments ($\beta$ = –0.046, p < .001). As in the main analysis, truth judgment rates and accuracy remained unaffected by mediation type.

\section{Discussion}
In our experiments, AI-mediated video communication substantially affected how people evaluated each other. Particularly in the strong AI-mediated treatment, where video subjects used synthetic avatars, participants' trust in the speaker and their confidence in their truth-lie judgments were reduced. While AI mediation changed the cues on which participants relied for their judgments, it did not affect how often participants suspected others of lying, nor did it improve or impair judgments accuracy. The observed decreases in trust and confidence were moderated by the type of environment, with larger decreases in mixed environments, where participants encountered a mix of original and AI-mediated videos. 

In the following, we discuss three possible interpretations of why AI-mediated video might undermine trust without triggering suspicion; how accuracy remains stable while cue reliance changes; and how confidence in judgment drops even though judgment accuracy does not change. We finish by outlining implications for design and policy.

\subsection{AI-Mediated Video Reduces Interpersonal Trust Without Raising Suspicion} 

AI-mediated video consistently reduced interpersonal trust (H1), particularly when video subjects were replaced by avatars in a mixed environment alongside more natural, unaltered video subjects (H5). The findings align with prior work on avatar-mediated communication in which low-fidelity representations are trusted less than natural faces \cite{pandaAllTogetherEffectAvatars2022,maNodsAgreementWebcamDriven2025}. 
The observed effect of reduced trust in strong AI mediation was amplified in mixed environments (Study 2), aligning with Expectancy Violations Theory (EVT), which posits that deviations from the internalized social expectations decrease trust in the speaker \cite{burgoonExpectancyViolationsTheory2015}. The mixed environment may have made AI mediation more salient and highlighted participants' expectations regarding how a speaker should present themselves on video, leading to more negative trust evaluations. Here, our findings align with and extend the Replicant Effect \cite{jakeschAIMediatedCommunicationHow2019}, in which, in a mix of human-generated and AI-generated content, the trustworthiness of subjects suspected of using generated content decreases as people begin to question each other's humanity. Our findings show that the Replicant Effect holds in the more dynamic medium of video and that even after people have become more accustomed to various forms of AI systems in recent years, AI-mediated communication still decreases interpersonal trust. Our work also highlights that trust is reduced even under weaker forms of AI mediation, such as retouching and virtual backgrounds.


Surprisingly, although participants found AI-mediated video subjects less trustworthy, they believed them just as often as they believed those in the original video. The stability in truth and lie rates is consistent with Levine's truth-default theory \cite{levineTruthDefaultTheoryTDT2014}, which posits that, by default, people believe others are telling the truth unless something triggers suspicion. 
While the visual unfamiliarity and synthetic nature of the avatars may have disrupted trust, it was not enough to trigger suspicion and override the default behavior of believing others. 


The apparent contradiction of reduced trust in avatars yet continued belief in them is best understood by distinguishing trust and belief as psychologically distinct processes with different implications \cite{holtonDecidingTrustComing1994}. While trust reflects an affective and relational judgment about another person's characteristics, belief is a cognitive judgment about whether a statement is factually true \cite{holtonDecidingTrustComing1994}. Our findings extend prior work on HCI and avatar-mediated communication \cite{pandaAllTogetherEffectAvatars2022, maNodsAgreementWebcamDriven2025} by showing that AI-mediated speakers can be believed without being trusted. That distinction matters because AI mediation can erode the interpersonal foundations of communication, such as trust, social connection and willingness to cooperate \cite{panImpactSelfavatarsTrust2017,schilkeTransparencyDilemmaHow2025}, without undermining social epistemology and without leading to a mediation environment in which people begin to question mediated statements.


\subsection{AI-Mediated Video Affects Cue Reliance But Not Deception Detection Accuracy}

The present work extends prior work on the effects of avatar-mediated communication \cite{pandaAllTogetherEffectAvatars2022} by examining their effect on lie detection. Here, our results largely align with well-documented findings from deception research \cite{depauloCuesDeception2003,bondAccuracyDeceptionJudgments2006}. In both studies, participants identified truths and lies with roughly 52–54\% accuracy, replicating the average truth-lie accuracy of 54\% reported in \cite{bondAccuracyDeceptionJudgments2006}. Participants were also more accurate at identifying truths than lies, an asymmetry known in deception research as the veracity effect \cite{levineAccuracyDetectingTruths1999}. 
However, accuracy rates did not differ meaningfully between the control condition, in which video subjects were unaltered, and the retouch or avatar conditions, in which facial expressions, eye contact and other visual behaviors were altered or removed. 

The stable accuracy across conditions, even under strong AI mediation, contradicts the predictions of cue-based deception detection theories \cite{ekmanLyingDeception1997,ekmanNonverbalLeakageClues1969}. These theories posit that deception is detected by leakage, that is, by observing a set of nonverbal cues involuntarily produced by the cognitive effort of lying. As avatar-mediated communication would substantially reduce the nonverbal cues that might give away a liar, we would expect accuracy to decrease in environments with reduced leakage. Our studies, however, show that accuracy neither decreased nor improved in the avatar condition. 

While we do not observe an effect of avatars on deception detection, the stable accuracy across conditions aligns with a common finding in lie detection research: people are not good at detecting lies \cite{depauloCuesDeception2003}, regardless of the type of mediation. Instead, the stable accuracy across conditions supports a heuristic view of lie detection \cite{bondAccuracyDeceptionJudgments2006,levineScientificEvidenceCue}, which holds that judgments are driven by content-based cues like plausibility, coherence and fact-checking, rather than leakage through involuntary nonverbal cues. These content-based cues, such as plausibility, coherence, and fact-checking, are arguably less affected by the use of retouching, backgrounds and video avatars.

While accuracy was not affected by AI mediation, participants shifted to rely on speech fluency, story consistency, and specificity, rather than to facial expressions or body language. This finding is again paralleled by the central tenet of Levine's truth-default theory \cite{levineTruthDefaultTheoryTDT2014}, which holds that content rather than demeanor is the basis for detecting deception. 
Based on Levine's truth-default theory, one might have expected the shift to content-based cues to lead to even greater accuracy, as when people pay more attention to content cues, they are more likely to notice inconsistencies. 
However, accuracy remained stable across conditions, which may be due to the nature of the interactions we studied: in our tasks, participants judged statements about unknown people, with no means to verify the claims or compare them with existing knowledge, limiting the potential to challenge the default assumption of truth. Although participants shifted to rely on content-based and verbal cues, without information to contextualize or fact-check the statements, participants may have rarely had the chance to notice inconsistencies that might have broken their truth-default bias.

\subsection{AI-Mediated Video Complicates Judgments and Reduces Judgment Confidence}

Participants' subjective confidence in their own judgments decreased in the mixed environment (Study 2), particularly for the avatar condition---again, despite their truth-lie rates and accuracy staying the same. 

The decrease in confidence in judgments for avatars aligns with prior findings on avatar-mediated communication \cite{pandaAllTogetherEffectAvatars2022} and is supported by both cue-based theories and Uncertainty Reduction Theory. While decades of deception research have shown that nonverbal cues like gaze or facial expressions are poor indicators of lying \cite{depauloCuesDeception2003}, people continue to rely on them, or at least, they believe they depend on them. When these familiar cues are removed or disrupted, people feel less confident about what to rely on instead. In this way, these cues serve a social-psychological function: they reduce uncertainty and help individuals feel familiar in social interactions. Uncertainty Reduction Theory \cite{bergerExplorationsInitialInteraction1975} posits that people are motivated to seek information that reduces ambiguity in social communication. AI-mediated communication interrupts that process by stripping away or synthetically simulating these social cues. This disruption holds even after people have gotten more accustomed to AI tools, and even for commonly used transformations such as lighting corrections and virtual backgrounds. Even if the removal of familiar cues does not impair accuracy, it leaves participants feeling less sure of themselves because they can't rely on some signals that they would usually rely on to make interpretation easier.



The most pronounced decline in judgment confidence is in Study 2, in which participants saw all three types of mediation within the same environment. The mix of mediation types likely heightened the salience of visual disruption and made avatars feel more unpredictable or "out of place". Here, Expectancy Violation Theory \cite{burgoonExpectancyViolationsTheory2015} posits that people have internalized expectations for how social interactions should look and feel, and when those expectations are violated, such as encountering a synthetic face after two natural ones, it creates friction in the interpretation of the communication. While friction does not make the person seem more deceptive, it may make the interaction more challenging to process. Previous research suggests that a disruption in processing fluency leads to lower confidence in judgments \cite{burgoonExpectancyViolationsTheory2015,zhouImpactItemDifficulty2023}. This interpretation aligns with participants' open-ended responses in our study, in which they describe avatars as "hard to read" or "off-putting", suggesting that a breakdown in fluency of interpretation makes people feel less confident about what to make of the communication.

\subsection{Implications}


Our findings show that AI-mediated video did not affect truth judgment rates or detection accuracy. As such, our findings challenge concerns that ``with their ability to alter users' appearances dramatically, beauty filters can facilitate deception''~\cite{marrPicturePerfectHidden}. Similarly, AI-mediated video has been discussed as a potential threat to people's ability to judge honesty accurately \cite{parkAIDeceptionSurvey2024d,WhyAIFilters}. Here, although we find that AI-mediated communication can affect trust and confidence, it does not substantially affect people's ability to detect lies in video communication. We note, however, that even under AI mediation, people's ability to detect deception remains close to chance.

However, consistent with prior work on trust in avatars \cite{pandaAllTogetherEffectAvatars2022}, we show that speakers using AI-mediated video were trusted less and participants felt less confident in their judgments \cite{depauloAccuracyConfidenceCorrelationDetection1997a,harveyConfidenceJudgment1997,oconnorModelsHumanBehaviour1989}. 
The mismatch between judgment performance and relational trust and confidence is an important implication \cite{bellemareSelfConfidenceReactionsSubjective2019,dautricheSubjectiveConfidenceInfluences2021} in contexts where the feeling of certainty of knowing can be as important as the decision itself. In high-stakes settings such as remote hiring, clinical evaluations or legal proceedings, lower confidence may lead to hesitation, increased caution or reduced assertiveness \cite{doubleConfidenceJudgmentsInterfere2024a,liuAsymmetricImpactDecisionmaking2024,patalanoInfluenceGroupDecision2011a,schoolerConfidenceMoralDecisionMaking2024}. 
Our results also broaden the debate about the risks of AI video beyond deepfakes \cite{hancockSocialImpactDeepfakes2021,popaDeepfakeTechnologyUnveiled2025}: even widely used filters like retouching and virtual backgrounds affect evaluations of trust and credibility in ways that matter \cite{afrooghTrustAIProgress2024,hancockAIMediatedCommunicationDefinition2020}. Our results show that the central risk of AI-mediated video may lie in the erosion of trust and confidence in judgments, particularly when mixed environments make mediation salient.

When designing video communication platforms, the question is how new AI-based features that may improve aesthetics and convenience might also undermine trust and confidence. Here, further research is needed to understand what elements are required for representational consistency within calls, what forms of disclosure might mitigate reductions in trust, and higher-fidelity or more expressive avatars may preserve the aspects of communication that people feel they need to feel confident in their judgments \cite{pandaAllTogetherEffectAvatars2022,maNodsAgreementWebcamDriven2025}. Our findings also highlight the need for more context-sensitive forms of AI mediation: communication tools could offer users different AI mediation levels depending on the situation—for example, realistic appearances in professional calls and more stylized options in informal chats—to adjust the communication to the expectations of the context and to calibrate the affordances of the communication based on context-specific needs.

\subsection{Limitations and Future Directions}

Participants evaluated short, prerecorded videos from \cite{lloydMiamiUniversityDeception2019} in which video subjects made simple personal statements about people unknown to the participants. Despite being incentivized with a bonus payment, the simulated scenario did not constitute a high-stakes situation for participants, and speakers in the video likely felt minimal pressure to lie. A low-stakes context may have muted the behavioral leakage or suspicion triggers on which cue-based theories depend. Future work should examine whether AI-mediated video has a different impact on high-stakes lies, where participants are more motivated to detect deception and speakers are under pressure. Such settings could include hiring, legal evaluations or sensitive interpersonal disclosures, where stakes and incentives are real and carry consequences. 

Furthermore, AI-mediated communication may be perceived differently in ongoing teams, family calls or workplace meetings. As the video subjects were strangers to the participants, the findings may not generalize to communication with more familiar groups, where prior work shows that familiarity can attenuate negative impressions of mediated cues \cite{mieczkowskiAIMediatedCommunicationLanguage2021} and that contextual knowledge can increase lie-detection accuracy. Future work could examine how AI mediation affects trust and confidence in contexts where familiarity and existing relationships shape impressions and social expectations.

In addition, the experimental setup lacks the fluid interactive nature of real-time video communication. In live conversations, participants can ask follow-up questions, interpret timing and adapt to social feedback. More research is needed to investigate how AI-mediated appearances affect interpersonal dynamics in live or semi-structured conversations, particularly in collaborative or conflict-prone contexts such as negotiations or interviews. 
Finally, although participants at the time of our study (August 2025) had substantial exposure to AI tools and weaker forms of AI-mediated video, such as retouching or virtual backgrounds, strong AI-mediated video is still unevenly adopted. As AI-mediated video and synthetic appearances become more normalized, user expectations and reactions may shift. Future work should track how perceptions of AI-mediated video evolve, including through longitudinal studies and cross-cultural comparisons, to assess whether increased exposure to AI reinforces or attenuates the observed effects.


\begin{acks}
We thank our research assistant, José Agostinho, for assisting the first author with processing the video stimuli and generating the avatars.

We acknowledge the use of ChatGPT for reviewing the author's original writing and for proposing phrasing improvements to increase clarity. All manuscript text was written and finalized by the authors.
\end{acks}

\bibliographystyle{ACM-Reference-Format}
\bibliography{references}

\appendix
\section{Appendix}
\subsection{Statistical models}
\label{models}

In the following, we provide the complete regression tables referenced in the results section, offering a comprehensive overview of all model specifications and coefficients presented in our analysis. Our statistical reporting draws on four different models: We fitted a linear mixed model (formula: outcome ~
AI mediation, estimated using REML and nloptwrap optimizer) to predict the outcome (reported trust, truth judgment rate, judgment accuracy and confidence) based on the type of AI mediation (original, retouches, or avatar-based) with a per-subject random fixed effect in the homogeneous environment in Study 1. The details are reported in Table \ref{table-model1}. We calculated an equivalent model for the mixed environment in Study 2 only, with statistics reported in Table \ref{table-model2}. We then fitted a linear mixed model across studies (estimated using REML and nloptwrap optimizer) to predict the outcome based on the interaction of the AI mediation and environment type (formula: outcome ~ environment + environment:ai-mediation) with a per-subject random fixed effect across both studies. The details are reported in Table \ref{table-model3}. Finally, we fitted an extended version of the previous model that also included covariates for participant age, gender, education, race, and experience with AI. The model details are reported in Table \ref{table-model4}. 
\appendix
\onecolumn

\begin{table}[t]
\centering
\caption{\textbf{Study 1} Linear mixed model predicting the outcome based on mediation type with a per-participant random fixed effect}
\label{table-model1}

\begin{tabular}{lcccc}
\toprule
 & Trust & Truth & Accuracy & Confidence \\
\midrule
(Intercept) & \num{0.510}*** & \num{0.599}*** & \num{0.523}*** & \num{0.564}*** \\
            & (\num{0.008})  & (\num{0.011})  & (\num{0.011})  & (\num{0.010})  \\
MediationRetouch & \num{-0.006} & \num{-0.017} & \num{-0.013} & \num{0.006} \\
            & (\num{0.012})  & (\num{0.016})  & (\num{0.016})  & (\num{0.014})  \\
MediationAvatar & \num{-0.025}* & \num{0.017} & \num{0.017} & \num{-0.012} \\
            & (\num{0.012})  & (\num{0.016})  & (\num{0.016})  & (\num{0.014})  \\
\midrule
SD (Intercept Subject) & \num{0.116} & \num{0.000} & \num{0.000} & \num{0.160} \\
SD (Observations)      & \num{0.235} & \num{0.490} & \num{0.499} & \num{0.197} \\
Num.Obs.               & \num{5802}  & \num{5802}  & \num{5802}  & \num{5802}  \\
R2 Marg.               & \num{0.002} & \num{0.001} & \num{0.001} & \num{0.001} \\
R2 Cond.               & \num{0.197} &  &  & \num{0.398} \\
AIC                    & \num{571.3} & \num{8219.8} & \num{8435.6} & \num{-808.1} \\
BIC                    & \num{604.6} & \num{8253.1} & \num{8469.0} & \num{-774.8} \\
ICC                    & \num{0.2}   &  &  & \num{0.4} \\
RMSE                   & \num{0.22}  & \num{0.49}  & \num{0.50}  & \num{0.18} \\
\bottomrule
\end{tabular}

\vspace{2pt}
{\footnotesize \emph{Note.} * p \num{< 0.05}, ** p \num{< 0.01}, *** p \num{< 0.001}.}
\end{table}

\begin{table}[t]
\centering
\caption{\textbf{Study 2} Linear mixed model predicting the outcome based on mediation type with a per-participant random fixed effect}
\label{table-model2}

\begin{tabular}{lcccc}
\toprule
 & Trust & Truth & Accuracy & Confidence \\
\midrule
(Intercept) & \num{0.499}*** & \num{0.597}*** & \num{0.541}*** & \num{0.575}*** \\
            & (\num{0.006})  & (\num{0.011})  & (\num{0.011})  & (\num{0.006})  \\
VideoTypeRetouch & \num{-0.021}** & \num{-0.021} & \num{-0.019} & \num{-0.019}** \\
            & (\num{0.008})  & (\num{0.015})  & (\num{0.015})  & (\num{0.006})  \\
VideoTypeAvatar & \num{-0.080}*** & \num{-0.020} & \num{-0.007} & \num{-0.046}*** \\
            & (\num{0.008})  & (\num{0.015})  & (\num{0.015})  & (\num{0.006})  \\
\midrule
SD (Intercept Subject) & \num{0.092} & \num{0.000} & \num{0.045} & \num{0.146} \\
SD (Observations)      & \num{0.246} & \num{0.493} & \num{0.497} & \num{0.207} \\
Num.Obs.               & \num{6240}  & \num{6240}  & \num{6240}  & \num{6240}  \\
R2 Marg.               & \num{0.016} & \num{0.000} & \num{0.000} & \num{0.005} \\
R2 Cond.               & \num{0.137} &  & \num{0.008} & \num{0.335} \\
AIC                    & \num{851.4} & \num{8912.1} & \num{9061.0} & \num{-482.1} \\
BIC                    & \num{885.1} & \num{8945.8} & \num{9094.7} & \num{-448.4} \\
ICC                    & \num{0.1}   &  & \num{0.0} & \num{0.3} \\
RMSE                   & \num{0.24}  & \num{0.49}  & \num{0.50}  & \num{0.19} \\
\bottomrule
\end{tabular}

\vspace{2pt}
{\footnotesize \emph{Note.} * p \num{< 0.05}, ** p \num{< 0.01}, *** p \num{< 0.001}.}
\end{table}

\begin{table}[t]
\centering
\caption{\textbf{Study 1 and 2} Linear mixed model predicting the outcome based on mediation and environment type with a per-participant random fixed effect}
\label{table-model3}

\begin{tabular}{lcccc}
\toprule
 & Trust & Truth & Accuracy & Confidence \\
\midrule
(Intercept) & \num{0.510}*** & \num{0.599}*** & \num{0.523}*** & \num{0.564}*** \\
            & (\num{0.008})  & (\num{0.011})  & (\num{0.011})  & (\num{0.009})  \\
ConditionRetouch & \num{-0.006} & \num{-0.017} & \num{-0.013} & \num{0.006} \\
            & (\num{0.011})  & (\num{0.016})  & (\num{0.016})  & (\num{0.014})  \\
ConditionAvatar & \num{-0.025}* & \num{0.017} & \num{0.017} & \num{-0.012} \\
            & (\num{0.011})  & (\num{0.016})  & (\num{0.016})  & (\num{0.014})  \\
ConditionMixed & \num{-0.011} & \num{-0.002} & \num{0.017} & \num{0.011} \\
            & (\num{0.010})  & (\num{0.015})  & (\num{0.016})  & (\num{0.011})  \\
ConditionMixed $\times$ VideoTypeRetouch & \num{-0.021}** & \num{-0.021} & \num{-0.019} & \num{-0.019}** \\
            & (\num{0.007})  & (\num{0.015})  & (\num{0.015})  & (\num{0.006})  \\
ConditionMixed $\times$ VideoTypeAvatar & \num{-0.080}*** & \num{-0.020} & \num{-0.007} & \num{-0.046}*** \\
            & (\num{0.007})  & (\num{0.015})  & (\num{0.015})  & (\num{0.006})  \\
\midrule
SD (Intercept Subject) & \num{0.104} & \num{0.000} & \num{0.000} & \num{0.153} \\
SD (Observations)      & \num{0.241} & \num{0.492} & \num{0.499} & \num{0.202} \\
Num.Obs.               & \num{12042} & \num{12042} & \num{12042} & \num{12042} \\
R2 Marg.               & \num{0.014} & \num{0.001} & \num{0.000} & \num{0.004} \\
R2 Cond.               & \num{0.169} &  &  & \num{0.366} \\
AIC                    & \num{1438.6} & \num{17128.1} & \num{17493.6} & \num{-1278.2} \\
BIC                    & \num{1497.8} & \num{17187.3} & \num{17552.8} & \num{-1219.1} \\
ICC                    & \num{0.2}   &  &  & \num{0.4} \\
RMSE                   & \num{0.23}  & \num{0.49}  & \num{0.50}  & \num{0.19} \\
\bottomrule
\end{tabular}

\vspace{2pt}
{\footnotesize \emph{Note.} * p \num{< 0.05}, ** p \num{< 0.01}, *** p \num{< 0.001}.}
\end{table}

\begin{table}[t]
\centering
\caption{\textbf{Robustness check} Study 1 and 2 Linear mixed model predicting each outcome from mediation and environment type with a per-participant random intercept and controls for demographic and experience covariates}
\label{table-model4}

\begin{tabular}{lcccc}
\toprule
 & Trust & Truth & Accuracy & Confidence \\
\midrule
(Intercept) & \num{0.347}*** & \num{0.546}*** & \num{0.565}*** & \num{0.445}*** \\
ConditionRetouch & \num{-0.006} & \num{-0.019} & \num{-0.011} & \num{0.003} \\
ConditionAvatar & \num{-0.027}* & \num{0.014} & \num{0.018} & \num{-0.020} \\
ConditionMixed & \num{-0.012} & \num{-0.002} & \num{0.019} & \num{0.009} \\
AgeNum & \num{0.000} & \num{0.001} & \num{-0.000} & \num{0.000} \\
GenderMale & \num{-0.005} & \num{-0.002} & \num{-0.003} & \num{0.024}** \\
GenderNon-binary & \num{-0.014} & \num{0.021} & \num{-0.004} & \num{-0.051} \\
EducationNum & \num{0.005}* & \num{0.008}* & \num{0.002} & \num{0.004} \\
RaceBlack or African American & \num{0.045}** & \num{0.001} & \num{0.013} & \num{0.074}*** \\
            & (\num{0.015}) & (\num{0.021}) & (\num{0.022}) & (\num{0.018}) \\
RaceIndigenous or Native & \num{0.074}* & \num{0.123}** & \num{-0.007} & \num{0.094}** \\
            & (\num{0.031}) & (\num{0.044}) & (\num{0.045}) & (\num{0.036}) \\
RaceMiddle Eastern or North African & \num{0.051} & \num{0.100} & \num{0.125} & \num{0.049} \\
            & (\num{0.063}) & (\num{0.092}) & (\num{0.093}) & (\num{0.075}) \\
RaceMultiracial or Mixed Race & \num{0.039}* & \num{-0.010} & \num{0.017} & \num{-0.011} \\
            & (\num{0.019}) & (\num{0.028}) & (\num{0.029}) & (\num{0.023}) \\
RacePacific Islander & \num{0.097} & \num{0.158} & \num{0.057} & \num{0.119} \\
            & (\num{0.081}) & (\num{0.117}) & (\num{0.119}) & (\num{0.096}) \\
RaceWhite or Caucasian & \num{0.058}*** & \num{0.015} & \num{0.025} & \num{0.044}** \\
            & (\num{0.013}) & (\num{0.019}) & (\num{0.020}) & (\num{0.016}) \\
EnglishLevelNum & \num{0.002} & \num{-0.071} & \num{-0.064} & \num{-0.068} \\
            & (\num{0.033}) & (\num{0.047}) & (\num{0.048}) & (\num{0.039}) \\
ExperienceVideoNum & \num{0.036}** & \num{-0.003} & \num{0.012} & \num{0.043}** \\
            & (\num{0.012}) & (\num{0.018}) & (\num{0.018}) & (\num{0.015}) \\
ExperienceAINum & \num{0.014} & \num{0.014} & \num{-0.000} & \num{0.075}*** \\
            & (\num{0.015}) & (\num{0.021}) & (\num{0.021}) & (\num{0.017}) \\
AIinteractionNum & \num{0.032}* & \num{-0.002} & \num{0.030} & \num{0.081}*** \\
            & (\num{0.014}) & (\num{0.020}) & (\num{0.020}) & (\num{0.016}) \\
AITrustNum & \num{0.090}*** & \num{0.104}*** & \num{-0.039} & \num{0.058}*** \\
            & (\num{0.014}) & (\num{0.021}) & (\num{0.021}) & (\num{0.017}) \\
ConditionMixed $\times$ VideoTypeRetouch & \num{-0.021}** & \num{-0.021} & \num{-0.019} & \num{-0.019}** \\
            & (\num{0.007}) & (\num{0.015}) & (\num{0.015}) & (\num{0.006}) \\
ConditionMixed $\times$ VideoTypeAvatar & \num{-0.080}*** & \num{-0.020} & \num{-0.007} & \num{-0.046}*** \\
            & (\num{0.007}) & (\num{0.015}) & (\num{0.015}) & (\num{0.006}) \\
\midrule
SD (Intercept Subject) & \num{0.098} & \num{0.000} & \num{0.000} & \num{0.142} \\
SD (Observations)      & \num{0.241} & \num{0.491} & \num{0.499} & \num{0.202} \\
Num.Obs.               & \num{12042} & \num{12042} & \num{12042} & \num{12042} \\
R2 Marg.               & \num{0.034} & \num{0.006} & \num{0.001} & \num{0.055} \\
R2 Cond.               & \num{0.171} &  &  & \num{0.368} \\
AIC                    & \num{1445.1} & \num{17202.4} & \num{17616.2} & \num{-1372.1} \\
BIC                    & \num{1630.0} & \num{17387.3} & \num{17801.1} & \num{-1187.2} \\
ICC                    & \num{0.1} &  &  & \num{0.3} \\
RMSE                   & \num{0.23} & \num{0.49} & \num{0.50} & \num{0.19} \\
\bottomrule
\end{tabular}

\vspace{2pt}
{\footnotesize \emph{Note.} * p \num{< 0.05}, ** p \num{< 0.01}, *** p \num{< 0.001}.}
\end{table}

\end{document}